\documentclass[11pt]{article}
\usepackage{graphicx}
\usepackage{amsmath}
\usepackage{amssymb}
\usepackage{amsxtra}
\usepackage{float}
\usepackage[T1]{fontenc}

\usepackage{subcaption, csquotes}

\usepackage[
	style=phys,
	biblabel=brackets,
	eprint=true,
	bibencoding=auto,
	backend=biber,
	sorting=none,
	maxbibnames=3,
	language=auto,
	autolang=other,
	doi=false,
	url=false
	]{biblatex}

\AtEveryBibitem{
	\clearfield{day}
	\clearfield{month}
	\clearfield{endday}
	\clearfield{endmonth}
}

\usepackage[colorlinks]{hyperref}

\usepackage{authblk}

\title{Soliton foam formation in the\\early Universe}
\author{A.~A.~Kirillov\thanks{\href{AAKirillov@mephi.ru}{AAKirillov@mephi.ru}}}
\author{B.~S.~Murygin\thanks{\href{MuryginBS@gmail.com}{MuryginBS@gmail.com}}}
\author{V.~V.~Nikulin\thanks{\href{VVNikulin@mephi.ru}{VVNikulin@mephi.ru}}}

\affil{
    National Research Nuclear University MEPhI
    \par
    (Moscow Engineering Physics Institute), 
    \par
    115409 Kashirskoe shosse 31, Moscow, Russia
    }

\date{}

\begin{document}

\maketitle

\begin{abstract}
    The formation of composite solitons produced by scalar fields without thermal phase transitions in the early Universe is considered. We present numerical simulations of the formation and evolution of soliton structures at the post-inflationary stage. The realistic initial conditions are obtained through the simulation of multiple quantum fluctuations during the inflation epoch. The initial field distributions allow to form local soliton clusters in the early Universe without the need for the thermal production of a soliton network throughout the Universe. We find that in three-dimensional space, the nontrivial composite field structures are formed in the form of <<soliton foam>>, consisting of closed domain walls, domain walls bounded by cosmic strings, and scalar field radiation. The possible cosmological implications of the soliton foam are discussed.
\end{abstract}



\section{Introduction}
\label{sec:intro}

Formation of topological defects in the early Universe has been widely discussed since the first papers \cite{1975ZhETF..67....3Z, 1976JPhA....9.1387K, 1980PhR....67..183K}. Their cosmological and astrophysical implications are of particular interest. Creation of strings and domain walls can be induced by a spontaneous symmetry breaking in a unified gauge theory. This leads to a soliton network formation which previously was considered as a natural source of large scale structure \cite{1976JPhA....9.1387K, 1985PhR...121..263V, 1989PhRvD..40..973A, 1989ApJ...347..590P, 1989CoNPP..19...25H, 1993NuPhB.406..452L, 1995RPPh...58..477H, 1998PhRvD..57.3317M}. In addition, a network was treated as a natural candidate for dark energy due to its negative equation of state parameter $w$ \cite{2003PhRvD..67d3519F}.
%and dark matter \cite{1999astro.ph..8047B, 1999PhRvD..60d3505B}. 
However, the abundance of modern observational data has greatly limited the available parameter space of the field models predicting solitons formation \cite{1998PhRvD..59b3508A, 2002PhR...364....1D, 2004APh....21..443C, 2005PhRvD..72b3513W, 2014PhRvL.112m1101A}. Actually, a domain wall network is ruled out by lot of observational data. On the other hand, a string network remains a viable model \cite{2015PhRvD..91d3001S, 2023JCAP...04..045H}.

However, the situation have to be revised if we consider inflationary soliton production mechanism \cite{2000hep.ph....5271R, 2001JETP...92..921R}. Usually a network is considered as the product of a thermal phase transition when the field effective potential acquires a non-trivial vacuum as the temperature decreases $V_\text{eff}(T)$. The temperature decrease due to cosmological expansion leads to solitons production as a network throughout the Universe. On the other hand, the solitons might be formed in the Universe locally in isolated clusters (opposite to the global network case) if its generation is determined by the multiple quantum fluctuations during the inflation. Quantum fluctuations of a scalar field during the inflation epoch could lead to suitable conditions allowing the local generation of solitons, which can produce primordial black holes (PBH) due to collapse of cosmic string loops \cite{1989PhLB..231..237H, 1991PhRvD..43.1106P, 1996PhRvD..53.3002C, 2018JCAP...11..008V} or domain wall bubbles \cite{2000hep.ph....5271R, 2001JETP...92..921R, 2020PhRvD.101b3513L}. Moreover, such black holes can form clusters and serve as nuclei for galaxies and early quasars \cite{2000hep.ph....5271R, 2001JETP...92..921R, 2005APh....23..265K, 2005GrCo...11...99D, 2008ARep...52..779D, 2019EPJC...79..246B}.

In this work, we consider the formation of solitons due to quantum fluctuations during inflation. We use realistic initial conditions simulated by the Starobinsky's inflation model \cite{1980PhLB...91...99S} and present the results of numerical 3D simulations of the dynamical evolution of formed soliton structures and discuss various effects of their evolution and its cosmological implications.

The rest of the paper is organized as follows. 
In sections~\ref{sec:setup}, \ref{sec:infcond}, \ref{sec:nummetods}, we describe our field model, generate inflationary initial conditions for fields, and define the numerical methods that we have used to simulate the soliton production and evolution. In section~\ref{sec:results}, we discuss the different effects that occur during the evolution of soliton structures and can be seen in the results of our simulations. Finally, we summarize and discuss some cosmological implications of our model.

\section{Non-thermal soliton formation}

In this work we study the non-thermal (inflationary) mechanism of soliton production in a scalar field. We do not consider any specific interaction of the scalar field with the plasma at the reheating stage and neglect changes in the shape of the effective field potential at finite temperature. To be specific, we work under the assumption that the temperature of the thermal phase transition (if there was one) is greater than the inflation scale $H_\text{inf}$ (that is, greater than the reheating temperature). Therefore, the effects of the thermal phase transition do not appear in the post-inflationary Universe.

In contrast to the thermal one, we consider the mechanism of soliton production due to inflationary quantum fluctuations. The mechanism is based on sub-inflationary scalar fields with squared mass $\mathcal{V}'' \ll H^2_\text{inf}$. Their classical evolution is <<frozen>> until the end of inflation, and the field potential is effectively flat compared to the inflation scale. The behaviour of such fields during the inflation period is determined by multiple quantum fluctuations, and after the end of inflation, the fields <<unfreeze>>, and their following evolution is determined by the classical field dynamics. The soliton production becomes possible if the initial fluctuations appear in a vicinity of a saddle point or a peak of the field potential \cite{2018JCAP...04..042G, 2021Physi...3..563M}. The goal of this paper is to simulate the classical dynamics of soliton formation after the fields <<unfreezing>> with inflationary initial conditions.

\subsection{Field model and potential}
\label{sec:setup}

In this paper, we focus on the formation of scalar field soliton structures immediately after the end of the inflation epoch. Let us take the model with two scalar fields
\begin{equation}
 \label{action}
  S = \int\sqrt{-g} \left[\frac{1}{2}(\partial\phi)^2 
   + \frac{1}{2}(\partial\chi)^2 
   - \mathcal{V}\left(\phi,\chi\right)\right] \mathrm{d}^4 x
   \,
\end{equation}
with the potential $\mathcal{V} ( \phi, \chi )$ of general form having at least one peak and one saddle point. Previously, it was shown that the soliton production is possible in this case \cite{2018JCAP...04..042G, 2020arXiv201107041K, 2021Physi...3..563M}. To be specific, let us take the form of the potential \cite{2021Physi...3..563M}
\begin{equation}
  \label{poten}
  \mathcal{V} ( \phi, \chi )
  = \frac{m^2}{2}\left(\phi^2+\chi^2\right)
  + \Lambda^4 \exp
  \left(
  - \dfrac{ ( \phi - \phi_0 )^2 + ( \chi - \chi_0 )^2 }{2\sigma^2}
  \right)
  \,,
\end{equation}
where $(\phi_0,\chi_0)$ are the coordinates of the local peak, which is a feature of this potential (it allows the formation of both strings and domain walls \cite{2021Physi...3..563M}). Further, for convenience, all parameters are expressed in the units of inflationary scale $H_\text{inf}$. The following parameters were chosen for simulation: $\Lambda = 0.2$, $\sigma = 1$, $m = 0.001$, $\phi_0 = 0$, $\chi_0 = 5$.

The field system evolves on the background of the expanding Universe with metric
\begin{equation}
  g_{\mu\nu}
  = \text{diag}
  \left[1,-a^2(t),-a^2(t),-a^2(t)\right]
  \,,
\end{equation}
where the scale factor $a(t)$ takes different expressions for different stages of the Universe evolution. For modeling purposes, we neglect the influence of the field system on the cosmological expansion.

In this work we are interested in the end of cosmological inflation as the start point for soliton formation. We use the approximate inflationary metric $a(t) = e^{Ht}$, where $H$ is the time-dependent Hubble parameter. Note, during the inflation $H = \mathrm{const}$, while at the exit of the inflation process, $H(t)$ decreases. The last one is the stage when a soliton foam is formed. Simulations of this period can also show most of the effects that occur at the scale of the soliton thickness. These effects include the emergence of strings and walls, their reconnecting, the formation of holes, and the emission of scalar field radiation. All these effects are discussed in more details in this article.

The energy density distribution of the field system \eqref{action} is found to be
\begin{equation}
 \label{edens}
  \varepsilon(\mathbf{x},t)
  = T^{00}
  = \frac12
  \left(
  \dot{\phi}^2
  + \dot{\chi}^2
  + \frac{(\nabla\phi)^2}{a^2}
  + \frac{(\nabla\chi)^2}{a^2}
  \right)
  + \mathcal{V}\left(\phi,\chi\right)
  \, 
  ,
\end{equation}
while the corresponding motion equations obey 
\begin{equation}
 \label{fieldeq}
  \begin{aligned}
    \ddot{\phi}
    + 3H\dot{\phi}
    - \frac{\nabla^2\phi}{a^2}
    =& - m^2 \phi
    + \frac{\Lambda^4}{\sigma^2} \left(\phi-\phi_0\right)
    e^{
      - \frac{\left(\phi-\phi_0\right)^2
      + \left(\chi-\chi_0\right)^2}{2\sigma^2}
      }
    \,,
    \\
    \ddot{\chi}
    + 3H\dot{\chi}
    - \frac{\nabla^2\phi}{a^2}
    =& - m^2 \chi
    + \frac{\Lambda^4}{\sigma^2} \left(\chi-\chi_0\right)
    e^{
      - \frac{\left(\phi-\phi_0\right)^2
      + \left(\chi-\chi_0\right)^2}{2\sigma^2}
      }
    \,.
  \end{aligned}
\end{equation}

To define the $H(t)$ time dependence at the end of inflation, let us take the Starobinsky's $f(R)$ inflationary model \cite{1980PhLB...91...99S}, which is in the best agreement with the last observational data on the cosmological inflation \cite{planck}. The Starobinsky's effective inflaton field $\xi$ obeys the equation
\begin{equation}
  \ddot\xi + 3H \dot\xi = - \mathcal{U}_\xi'
  \, ,
  \label{eq:EqM.inflaton}
\end{equation}
where the potential has the form
\begin{equation}\label{inf_pot}
  \mathcal{U}(\xi)
  = \frac{3 M^2}{32 \pi^2}
  \left(1 - e^{-\sqrt{16 \pi\,/3}\,\xi}\right)^2
  .
\end{equation}
The inflaton field $\xi$ is expressed in units of $[M_\text{pl}]$. Here, $M$ is the parameter of the quadratic $f(R)$-gravity (expressed in units of $[H_\text{inf}]$), and the ratio between the inflation energy scale $H_\text{inf}$ and the Planck scale $M_\text{pl}$ depends on $M$. In this paper, the value $M$ is chosen in such a way as to ensure inflationary energy scale $H_\text{inf}=10^{13}$~GeV, which is consistent with the cosmic microwave background (CMB) observational data \cite{planck}. The Hubble parameter at the inflation stage:
\begin{equation}\label{hubble}
  H(t)
  = \sqrt{\frac{8 \pi}{3}
    \left(\frac{1}{2}\dot{\xi}^2 +\mathcal{U}(\xi)\right)}
  \,.
\end{equation}
The time-dependent Hubble parameter $H(t)$ is expressed in units of $[H_\text{inf}]$, corresponding to parameter $H$ at the beginning of the inflation.

\subsection{Inflationary initial conditions}
\label{sec:infcond}

% In this article, we explored not only the specific initial conditions that lead to soliton production or certain soliton effects, but also attempted to simulate more general and realistic initial conditions under which the field system $(\phi, \chi)$ starts the evolution in the early Universe. 

% In this paper, we consider the non-thermal inflationary mechanism for the formation of solitons. The mechanism is based on sub-inflationary scalar fields \eqref{fieldeq} with squared mass $\mathcal{V}''(\phi,\chi) \ll H^2_\text{inf}$, the classical evolution of which is <<frozen>> (due to the second term in \eqref{fieldeq}) until the end of inflation. The behaviour of such fields during the inflation period is determined by multiple quantum fluctuations, and after the end of inflation, the fields <<unfreeze>>, and their following evolution is determined by the classical field dynamics. The soliton production becomes possible if the initial fluctuations appear in a vicinity of a saddle point or a local peak of the field potential \cite{2018JCAP...04..042G, 2021Physi...3..563M}. Our goal is to simulate the classical dynamics of soliton formation after the fields <<unfreezing>>. In this section, we discuss and model inflationary fluctuations of the scalar fields \eqref{fieldeq}, which determine the initial conditions for our simulation.

In this section, we discuss and model inflationary fluctuations of the scalar fields \eqref{fieldeq}, which determine the initial conditions for our simulation. It is known that the statistics of quantum fields is determined by the two-point correlation function $\langle 0| \phi(\mathbf{x}) \phi(\mathbf{y}) |0 \rangle$. This fact (which is a consequence of Wick's theorem) makes it possible to use the apparatus of Gaussian random fields to model <<frozen>> quantum fluctuations \cite{1982PhLB..116..335L, 2005hep.th....3203L}. The Fourier image of the two-point correlator (for isotropic and homogeneous statistics) defines the power spectrum of quantum fluctuations
\begin{equation}
  \langle 0| \phi(\mathbf{k}) \phi(\mathbf{k}') |0 \rangle = \frac{P(k)}{(2\pi)^3}\,\delta(\mathbf{k}+\mathbf{k}') \, .
\end{equation}
The power spectrum of inflation-generated fluctuations is known 
\begin{equation}
  P(k) = k^{-d + (n_s - 1)} , 
\end{equation}
where $d=3$ is the number of spatial dimensions and $(n_s - 1)$ defines the spectrum tilt. We have no reason to assume that the spectrum tilt for this scalar field be significant in our Universe (for example, the tilt of the curvature perturbations \cite{planck} is very small and generated by inflaton scalar fluctuations), so for the limited purposes of our simulation, we can consider $(n_s - 1)\approx 0$.
% The Gaussian property of cosmological fluctuations is also confirmed by the CMB observations.

To obtain a picture of <<frozen>> quantum fluctuations we simulate the random Gaussian scalar fields $\phi_{P}(\mathbf{x})$, $\chi_{P}(\mathbf{x})$ with the desired power spectrum $P(k)$. The following known technique is used. First, we obtain the Fourier image of the random Gaussian field:
\begin{equation}
  % \label{gaussrf}
  \phi_P(\mathbf{k}) = \sqrt{P(k)}\,\phi_\text{w}(\mathbf{k})   \, , 
  \quad \text{with} \quad 
  \phi_\text{w}(\mathbf{k}) = \mathcal{F}\{\phi_\text{w}(\mathbf{x})\} \, ,
\end{equation}
% \begin{equation}\label{gaussrf}
%   \begin{aligned}
%     \phi_P(\mathbf{x}) &= \int \phi_P(\mathbf{k})\,e^{-i\mathbf{k}\mathbf{x}}\,\frac{\mathrm{d}^3 k}{\left(2\pi\right)^{3/2}}
%     \,,\\
%     \phi_P(\mathbf{k}) &= \sqrt{P(k)}\,\phi_\text{w}(\mathbf{k})
%     \,,\\
%     \phi_\text{w}(\mathbf{k}) &= \int \phi_\text{w}(\mathbf{x})\,e^{i\mathbf{k}\mathbf{x}}\,\frac{\mathrm{d}^3 k}{\left(2\pi\right)^{3/2}}
%     \,,
%   \end{aligned} 
% \end{equation}
where $\mathcal{F}$ is the Fourier transform and $\phi_\text{w}(\mathbf{x})$ is the randomly generated spatial white noise (uncorrelated pixels with the standard normal distribution). From the Fourier image we restore the Gaussian random field
\begin{equation}
\phi_P(\mathbf{x}) = \mathcal{F}^{-1}\{\phi_P(\mathbf{k})\}\,.
\end{equation}
The same procedure has been done for $\chi_P(\mathbf{x})$. All our calculations use the fast Fourier transform algorithm (FFT).

To get a realistic picture of inflationary quantum fluctuations (our initial conditions), we need to set the required field dispersion and average by rescaling $\phi_P(\mathbf{x})$:

\begin{equation}\label{gaussrf}
  \phi(\mathbf{x}) 
   = \phi_P(\mathbf{x}) \, \dfrac{\sigma[\phi(\mathbf{x})]}{\sigma[\phi_P(\mathbf{x})]} + \langle\phi(\mathbf{x})\rangle \, ,
\end{equation}
where $\sigma[\phi(\mathbf{x})]$ is the required field dispersion, determined by the duration of quantum fluctuations (defined below), while the dispersion of the standard normally distributed field $\phi_P(\mathbf{x})$ is $\sigma[\phi_P(\mathbf{x})] = 1$. The last term $\langle\phi(\mathbf{x})\rangle$ is the required average.

% \begin{enumerate}
%   \item 
%    For the required power spectrum $P(k)$, the momentum-space image of the scalar fields is calculated $\hat{\Phi}_P(k) = P(k)\, \hat{\Phi}_\text {w}(k)$. Here, $\hat{\Phi}_\text {w}(k)$ is the Fourier image of the random pre-generated white noise fields $\Phi_\text{w}(x) = \big( \phi_\text{w}(x), \chi_\text{w}(x) \big)$ in the coordinate space (the pixel values of these fields are uncorrelated and have the standard normal distribution).

%   \item 
%    By the inverse Fourier transformation of calculated $\hat{\Phi}_P(k)$, we obtain the scalar field $\Phi_P(x) = \big( \phi_\text{P}(x), \chi_\text{P}(x) \big)$ in the coordinate space with the required power spectrum. All calculations use the fast Fourier transform algorithm.

%   \item 
%    Next, the field values are normalized and shifted to achieve the required dispersion $\sigma[\Phi_P(x)]$ and the average value $\langle\Phi_P(x)\rangle$.
% \end{enumerate}

The average $\langle\phi(\mathbf{x})\rangle$ is determined by the arbitrary initial field values $\phi_\text{in}$ from which the simulated region of space is formed during the inflation (at the moment when this region was causally connected, i.e. had a horizon size $H^{-1}_\text{inf}$). These initial values influence the quantity and type of solitons that will arise as a result of the field evolution (we model and discuss different cases the section \ref{sec:foam}).

The dispersion $\sigma[\phi(\mathbf{x})]$ is determined by the number of e-folds $\Delta N = \ln (L/H^{-1}_\text{inf})$ that were required for the inflationary expansion from the horizon size $H^{-1}_\text{inf}$ to the simulated size $L$. The dispersion per one e-fold for effectively massless ($\mathcal{V}''(\phi,\chi) \ll H^2_\text{inf}$) scalar field is well known $\sigma_e = H_\text{inf}/2\pi$ \cite{2005hep.th....3203L}, and allows us to calculate $\sigma[\phi(\mathbf{x})] = \sigma_e\sqrt{\Delta N} = \sigma_e\ln (L/H^{-1}_\text{inf})$.

\begin{figure}[H]
  \centering
  \includegraphics[width=\linewidth]{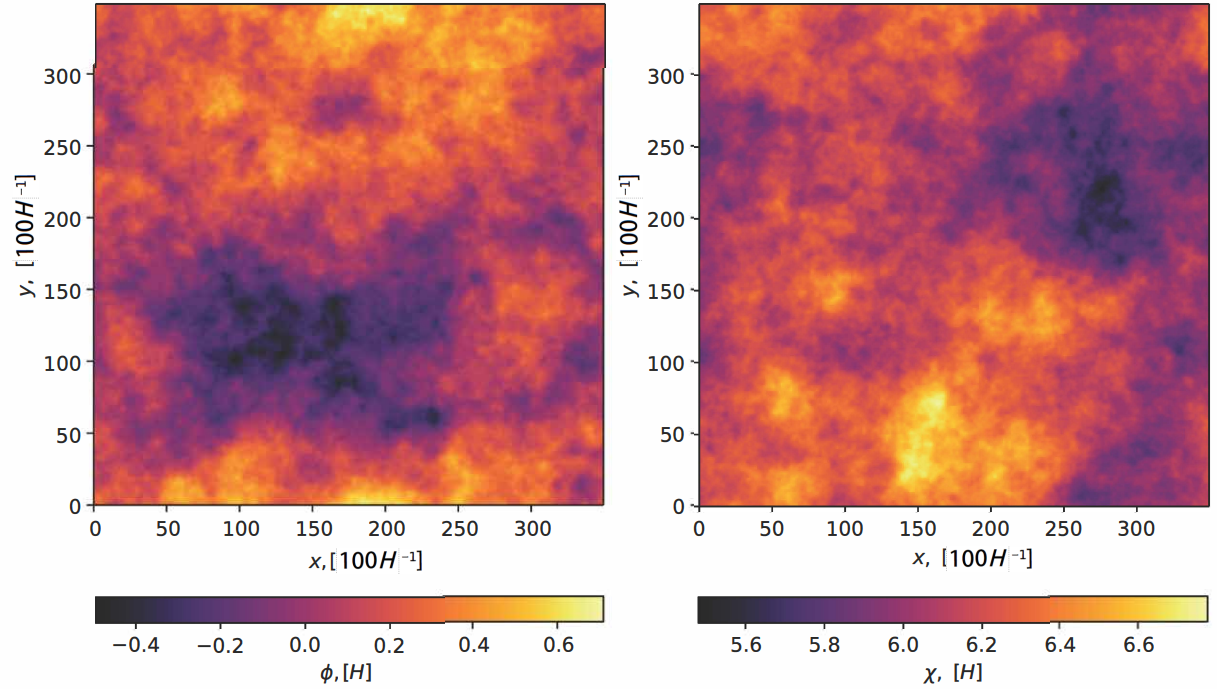}
  \caption{
    The result of 3D simulation of the initial distributions produced by quantum fluctuations in scalar fields $\phi(\mathbf{x})$, $\chi(\mathbf{x})$ during inflationary epoch. The xy-slice of 3D simulation is shown.
   % The initial scalar field conditions resulting from quantum fluctuations during inflationary epoch ($xy$-slice of 3D cube is taken). 
   All field values and distances are expressed in Hubble units $[H_\text{inf} = 10^{13} \, \text{GeV}]$.
   The simulation size $L^3=350^3\cdot100^3\,[H_\text{inf}^{-3}]$, initial field values $\left(\phi_\text{in} = 0,\,\chi_\text{in} = 6\right)$, the power spectrum $P(k)\approx k^{-3}$ \cite{planck}.
  }
  \label{fig:init_fields}
\end{figure}

Our 3D-simulation of the scalar fields distribution produced by inflationary quantum fluctuations is shown in Figure~\ref{fig:init_fields}. The resulting pair of fields $\phi(\mathbf{x})$, $\chi(\mathbf{x})$ remain in this state until they are <<unfrozen>> during the termination of the cosmological inflation (when $H^2(t) \lesssim \mathcal{V}''(\phi,\chi)$). This moment determines the beginning of our finite difference numerical simulation of the fields $\phi(\mathbf{x},t)$, $\chi(\mathbf{x},t)$.

Due to limited computing resources, the size of the simulation does not allow us to capture any cosmologically large regions of space and periods of time. However, this is not required to study the very dynamics of the soliton foam formation which occurs on the spatial scale of soliton thickness and the time scale of the <<unfreezing>> of field system at the end of inflation. A simulation size 1--2 orders of magnitude larger than these scales should already be sufficient, and it corresponds to the region of space generated during the last 4--5 e-folds of inflation (with the initial conditions (Figure~\ref{fig:init_fields}) preceding the <<unfreezing>>).

\section{Numerical methods}
\label{sec:nummetods}

For the numerical simulations, we use the explicit finite difference method (FD). Simulations were carried out using the linear algebra Python libraries \texttt{numpy}, \texttt{cupy} and \texttt{cupyx} which are optimized for CUDA framework for computing on NVidia Graphics Processors (GPU). The field values are set on the cubic 3D grid of size $L^3$ at three consecutive time layers. The spatial step is chosen as $\Delta x = 1$ in units of the selected scale $[H^{-1}_\text{inf}]$, and the time step is chosen as $\Delta t = 1/2$ in accordance with the Courant condition to ensure the numerical stability of solutions. We have used the numerical schemes with the accuracies $O(\Delta t^2 + \Delta x^2)$.

First, we need to understand how the inflaton field behaves when exiting the inflationary process. To solve the inflaton field equation \eqref{eq:EqM.inflaton}, we use the 2-d order Runge–Kutta method. We consider the inflaton field as a uniform background field. The Hubble parameter $H(t)$ is calculated according to~\eqref{hubble} (and hence the scale factor $a(t) \equiv e^{H(t) t}$). The initial value $H(0)$ is chosen so as to simulate the very end of the inflation (the last e-folds). It was during this stages the initial field conditions (Fig.~\ref{fig:init_fields}) <<unfreeze>> and the evolution of the fields begins. The evolution of the main fields \eqref{fieldeq} is calculated using the 2-d order difference scheme taking into account the coordinate dependence of the fields.

The initial field distributions $\phi(\mathbf{x},0)\,, \chi(\mathbf{x},0)$ for simulation are generated with the procedure described in section~\ref{sec:infcond}. We use the cyclic boundary conditions (both for the numerical scheme and for the initial conditions): the field values and derivatives are identified at opposite boundaries of the simulated volume. Thus, the simulated volume $L^3$ effectively has the topology of a torus, which is acceptable as long as we study phenomena that are much smaller than $L$. To visualize the result, we periodically calculate the energy density distribution \eqref{edens} using 2-d order difference scheme.

The computations are arithmetically intensive due to the complex expression of the potential \eqref{poten} as well as due to the number of intermediate calculation stages on each FD time step. To speed up such type of computations, the technique of <<fusing>> CUDA-cores is used (storing intermediate results in the GPU-cache). To increase the arithmetic intensity, the special <<tiling>> implementations for FD stencils were used \cite{sai2020accelerating, cai2018extreme}.

Note, the known problem of rapid decrease in wall thickness in the comoving coordinates (decrease to the grid scale) does not appear in our results due to the small time window of the simulation, which is sufficient to study the dynamics of soliton formation and evolution at the given spatial scales, but not sufficient to seriously increase the post-inflationary $a(t)$ factor.

\section{Results}
\label{sec:results}

In our numerical simulations, we find a complex soliton structure, which we called <<foam>>, consisting of different types of solitons and field radiation. Solitons that form this foam can be of two types: domain walls (forming closed bubbles) and strings attached to the domain walls (forming holes).

The types of emerging topological defects are uniquely determined by the topology of the set of equivalent vacua of theory. In the considered theory \eqref{poten}, there are two characteristic scales: $m$ and $\Lambda$. At energies $\sim \Lambda \gg m$ the potential has approximately U(1)-symmetry due to a Gaussian-like peak. The set of approximately equivalent vacua has a circle topology, which should lead to the appearance of string-type defects when U(1)-symmetry is spontaneously broken \cite{vilenkin}.

On energy scales $\sim m$, the vacua from the previously described circle cease to be equivalent and only one exact vacuum remains (up to a revolution of $2\pi n,\ n\in \mathbb{Z}$ around the peak of the potential \eqref{poten}). The latter leads to the appearance of wall-like defects: jumps in energy density when passing through a saddle point on the opposite side of the peak. In such a model, domain walls can be either isolated (in which case they should be closed into a bubble) or attached to strings (for which the closed string will be the boundary). At the same time, due to the possibility of $|n| > 1$, multilayer domain walls are possible \cite{2018JCAP...04..042G}. However, we do not observe such objects in our simulation due to the extremely low probability of their occurrence.

Such configurations was initially predicted for soliton networks \cite{1982PhRvD..26..435K, 1985PhR...121..263V}; however, as it is shown in \cite{2021Physi...3..563M} and in this article, such soliton structures remain valid for the inflationary soliton formation mechanism.
In our case, solitons of both types arise when inflation ends and the characteristic energy $H$ drops. During this stepwise process, first $H \sim \Lambda \gg m$ occurs with the formation of strings, and then $H \sim m$ occurs with the formation of attached or closed domain walls. These two types of soliton structures evolve differently, the various effects of their evolution are discussed below. Due to the relative proximity of the $\Lambda$ and $m$ scales, the formation of both types of solitons occurs quickly, and we will be interested in its final result --- a complex <<soliton foam>>.

The figures in this section show the spatial energy density distributions of field system. For better clarity, the individual effects are demonstrated in 2D plots (which can be viewed as a flat slice of a 3D simulation). As we will show below, these individual effects appear in full 3D simulations as part of complex soliton structures. For greater clarity, the first frames of the figures were taken after the simulation time $t \sim 10^2 \, H_\text{inf}^{-1}$ (so that the small-scale field inhomogeneities that blur the picture had time to relax). As expected, the characteristic size of the inhomogeneities is of the order of the horizon size at the simulation moment $t$.

As mentioned in the previous section, the scale of the simulations does not allow us to make cosmological predictions, but it does allow us to study all the effects that arise in the soliton foam. Moreover, the information about the density distribution of the various soliton foam components, obtained in these limited-scale simulations, can be used for further cosmological studies due to the scale invariance of the inflation process.

\subsection{Soliton interconnection}

The interconnection effect for cosmic strings was discussed in \cite{1995RPPh...58..477H, 1998PhRvD..57.3317M} and references within. If a string intersects itself, the formation of closed loops is possible, which could lead to the production of black holes due to the subsequent collapse of these loops \cite{1989PhLB..231..237H, 1991PhRvD..43.1106P, 1996PhRvD..53.3002C, 2018JCAP...11..008V}. Walls interconnection is discussed in \cite{1989ApJ...347..590P}. 

\begin{figure}[H]
  \centering
  \includegraphics[width=1.0\linewidth]{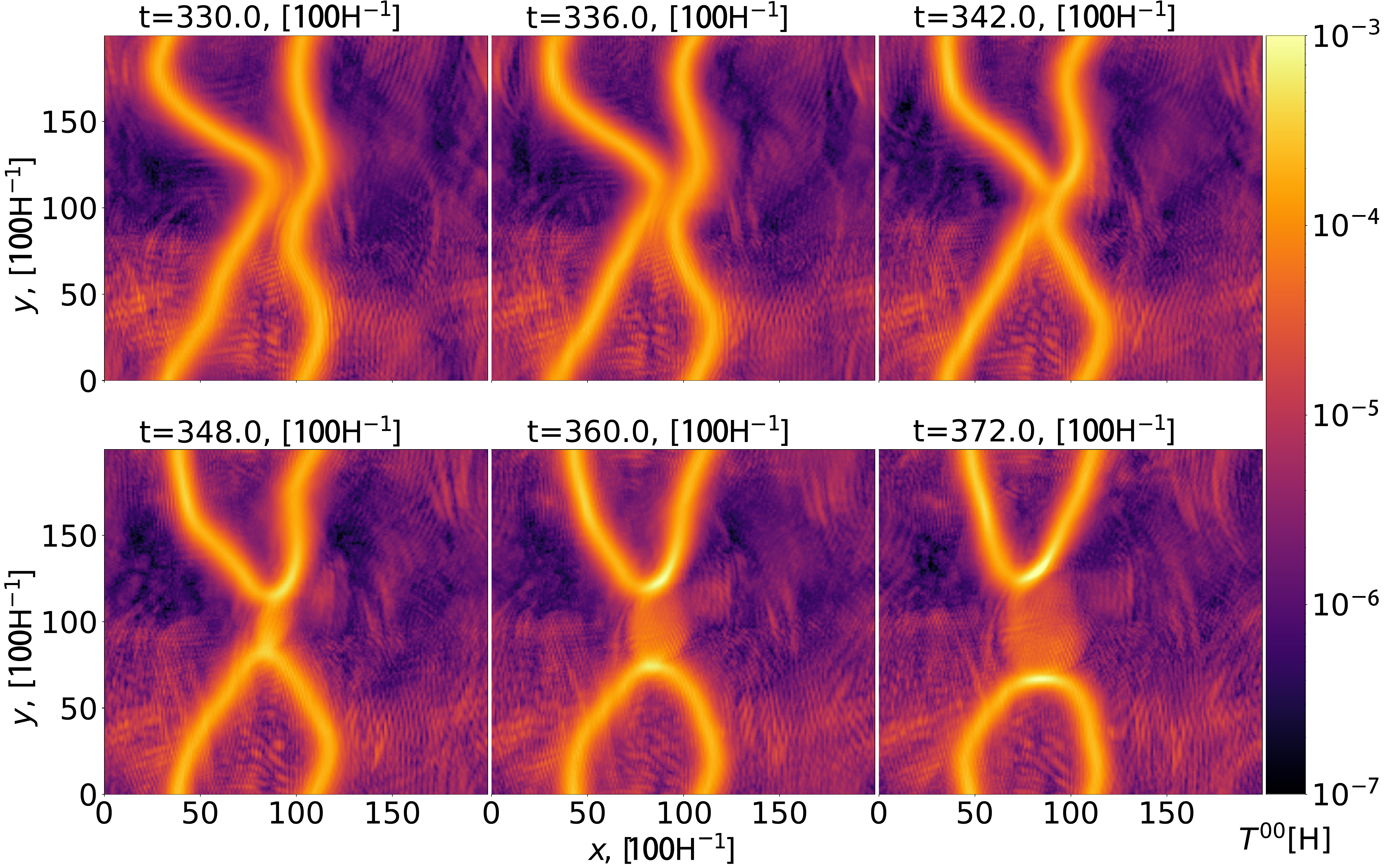}
  \caption{Interconnection of domain walls. The xy-slice of the 3D simulation is shown. The color scale represents the spatial energy density $\varepsilon\equiv T^{00}$ \eqref{edens} of the field system $\left(\phi,\chi\right)$. The field initial conditions (Fig.~\ref{fig:init_fields}) are simulated to produce pair of close domain walls. When domain walls get close to each other at a distance of the order of their thickness (top row), energy density can redistribute, resulting in their reconnection (bottom row). Energy density, distances and time are expressed in Hubble units $[H_\text{inf} = 10^{13}\,\text{GeV}]$.
  }
  \label{fig:interconnect}
\end{figure}

In our case, the interconnection of solitons comes naturally from the direct simulation of the field system. As a result of mutual intersection of solitons, they can merge, which is shown in Figure~\ref{fig:interconnect}. A similar effect can also occur due to self-intersection in a single soliton. In this case, instead of merging two solitons into one, there will be a process where a single soliton decays into two separate solitons. For instance, a string could produce a closed loop, and a domain wall could produce a bubble in the case of self-intersection.

\subsection{Spontaneous formation of closed strings}\label{sec:holes}

\begin{figure}
    \centering
  \includegraphics[width=0.9\linewidth]{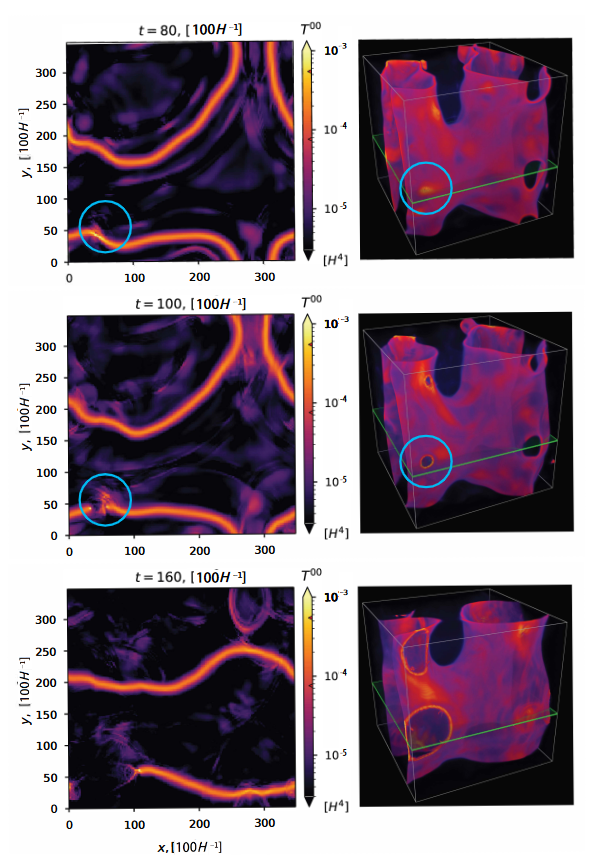}
  \caption{
   Formation of hole in the domain wall. The color scale represents the energy density $\varepsilon\equiv T^{00}$ \eqref{edens} of field system $( \phi, \chi )$. The pictures on the right show the 3D energy density distributions, while the pictures on the left show the 2D slices (green planes). In the regions with the increased energy density (marked by the blue circle), the wall ruptures with the formation of growing hole surrounded by the closed string. The formation of hole is accompanied by radiation emission. Energy density, distances and time are expressed in Hubble units $[H_\text{inf} = 10^{13}\,\text{GeV}]$.
  }
  \label{fig:tear}
\end{figure}

If a wall surface is significantly curved, holes may form on it. The edges of these holes become closed strings \cite{2021Physi...3..563M}. Thus, as a result of classical evolution of a curved domain wall, a domain wall bounded by a closed string can be formed. As shown in Figure~\ref{fig:tear}, the formation of several holes within one domain wall is possible. It is worth noting that the formation of holes in the walls is also possible through tunneling, although the probability of this process is low \cite{1985PhR...121..263V}.

Domain walls bounded by strings can appear without the need for dynamical field evolution, as it takes place in our case. They can form right after the beginning of postinflationary evolution due to specific initial conditions. However, despite the different formation mechanisms, the evolution of such structures follows the same pattern: holes bounded by string expand due to the tension of the domain wall leading to gradual decay of the wall.

The expansion of holes in domain walls bounded by strings emits a significant amount of scalar field radiation which can be interpreted (far from the decaying domain wall) as dark matter particles with the mass $m$. We discuss it in more detail in section~\ref{sec:discuss}.

Note, the topological possibility of hole formation is the consequence of the uniqueness of the vacuum in our model \eqref{poten}. However, in a multiple-vacuum model, the formation of holes is possible within those domain walls that separate regions with only one vacuum value if there is a peak in this region. While in the case of domain wall separating regions with different vacuum values, the formation of holes is topologically impossible.
% \begin{figure}[H]
%   \centering
%   \stackinset{l}{0.61\linewidth}{b}{0.11\linewidth}
%   {\tikz{\draw[cyan, thick] circle (0.04\linewidth)}}
%   {\stackinset{l}{0.11\linewidth}{b}{0.05\linewidth}
%   {\tikz{\draw[cyan, thick] circle (0.04\linewidth)}}
%   {\includegraphics[width=0.9\linewidth]{hole-1.pdf}}}

%   \stackinset{l}{0.61\linewidth}{b}{0.11\linewidth}
%   {\tikz{\draw[cyan, thick] circle (0.04\linewidth)}}
%   {\stackinset{l}{0.11\linewidth}{b}{0.05\linewidth}
%   {\tikz{\draw[cyan, thick] circle (0.04\linewidth)}}
%   {\includegraphics[width=0.9\linewidth]{hole-2.pdf}}}

%   \includegraphics[width=0.9\linewidth]{hole-3.pdf}
%   \caption{
%    Formation of hole in the domain wall. The color scale represents the energy density $\varepsilon\equiv T^{00}$ \eqref{edens} of field system $( \phi, \chi )$. The pictures on the right show the 3D energy density distributions, while the pictures on the left show the 2D slices (green planes). In the regions with the increased energy density (marked by the blue circle), the wall ruptures with the formation of growing hole surrounded by the closed string. The formation of hole is accompanied by radiation emission. Energy density, distances and time are expressed in Hubble units $[H_\text{inf} = 10^{13}\,\text{GeV}]$.
%   }
%   \label{fig:tear}
% \end{figure}

\clearpage

\subsection{Radiation emission}
\label{sec:radiation}

\begin{figure}[H]
  \centering
  \includegraphics[width=1.0\linewidth]{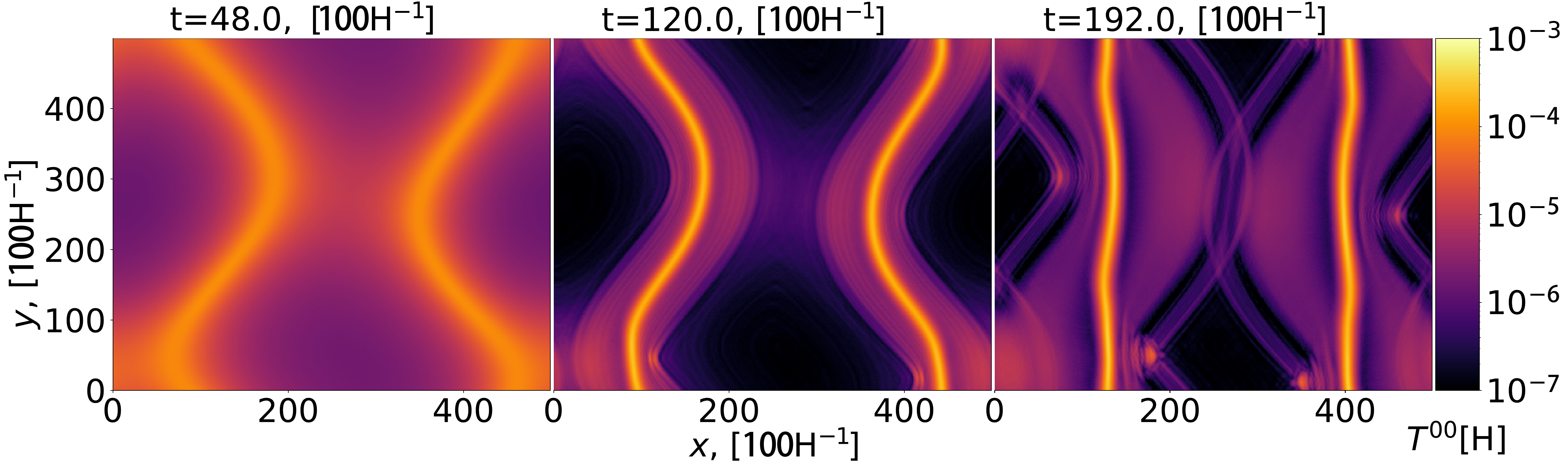}
  \caption{
   Radiation emission through the relaxation of wall curvature. The xy-slice of the 3D simulation is shown. The color scale represents the energy density $\varepsilon\equiv T^{00}$ \eqref{edens} of the field system $( \phi, \chi )$. The separated wave propagates to the spatial infinity, and can be interpreted as a flow of scalar particles of the field system. Energy density, distances and time are expressed in Hubble units $[H_\text{inf} = 10^{13}\,\text{GeV}]$.
  }
  \label{fig:rad_emission}
\end{figure}

Oscillations of solitons can be a source of radiation, which can be interpreted as particles (asymptotically far from the soliton). This effect is considered in the literature both for strings and domain walls \cite{1984PhRvD..30.2046V, 1987NuPhB.293..812B, 1987PhLB..189..397S, 1995RPPh...58..477H, 2023JCAP...04..045H}. Due to inflationary generation (by quantum <<random walk>>), solitons have significant surface curvature. During evolution, solitons tend to get rid of these curvatures. The release of excess surface energy occurs due to the emission of a scalar field radiation (Figure~\ref{fig:rad_emission}).

\begin{figure}[H]
  \centering
  \includegraphics[width=1.0\linewidth]{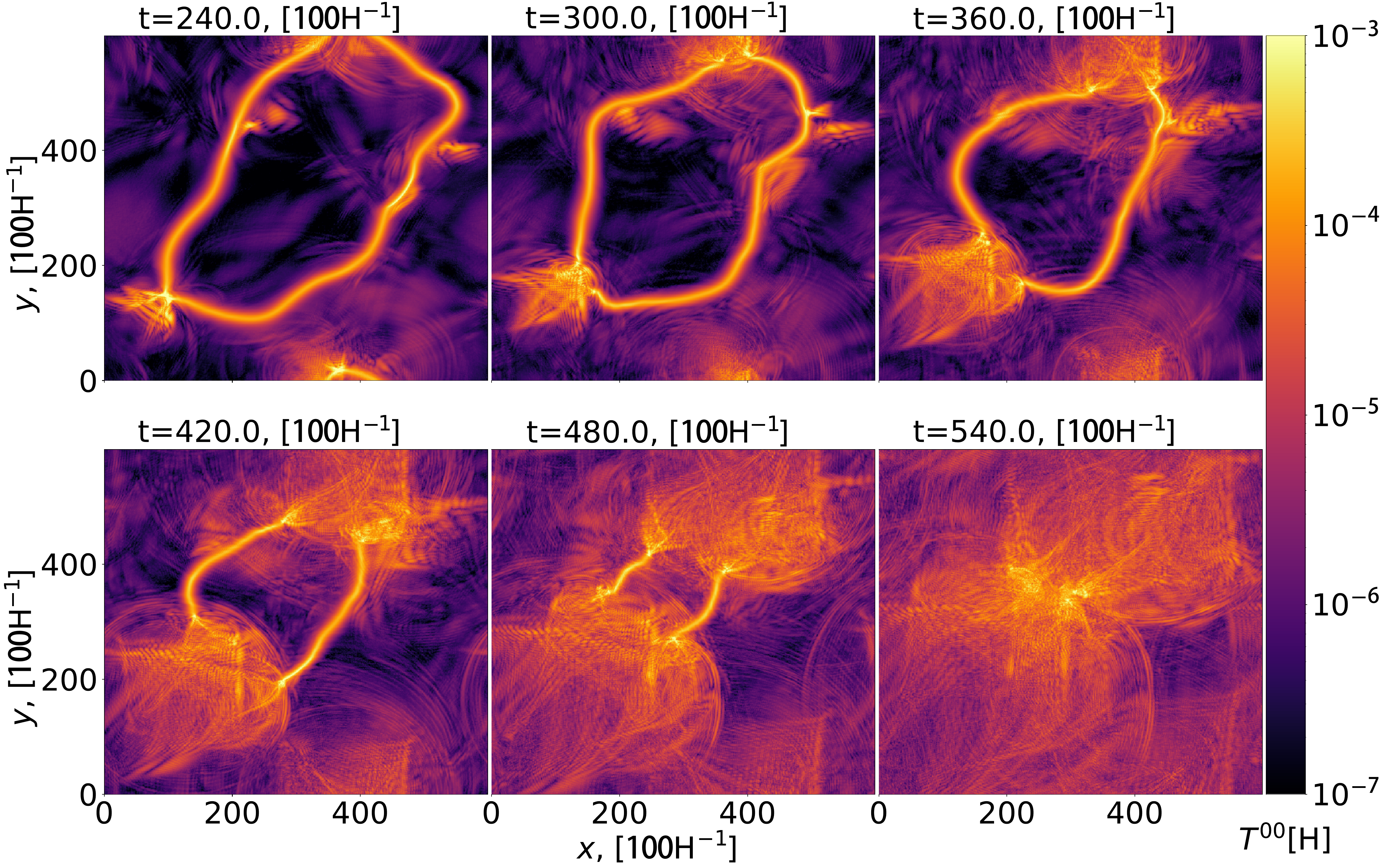}
  \caption{
   Radiation emission through the wall rupture. The xy-slice of the 3D simulation is shown. The color scale represents the energy density $\varepsilon\equiv T^{00}$ \eqref{edens} of the field system $( \phi, \chi )$. This process is similar to that shown in Figure~\ref{fig:tear} (this Figure shows 2D slice). Two holes are formed in the domain wall leading to the shrinking of the domain wall with the intense emission of radiation. Energy density, distances and time are expressed in Hubble units $[H_\text{inf} = 10^{13}\,\text{GeV}]$.
  }
  \label{fig:tear_2}
\end{figure}

In classical field simulations, the radiation emission appears as wave packet emission. In addition, observe the expected effects of wave packets interaction and reflection/refraction when they pass through domain walls. The radiation emission can also occurs as a result of the formation of holes in domain walls, which leads to contraction of the entire domain wall structure (Figure~\ref{fig:tear_2}).

\subsection{Relaxation of curvature and collapse of walls}

\begin{figure}[H]
  \centering
  \includegraphics[width=1.0\linewidth]{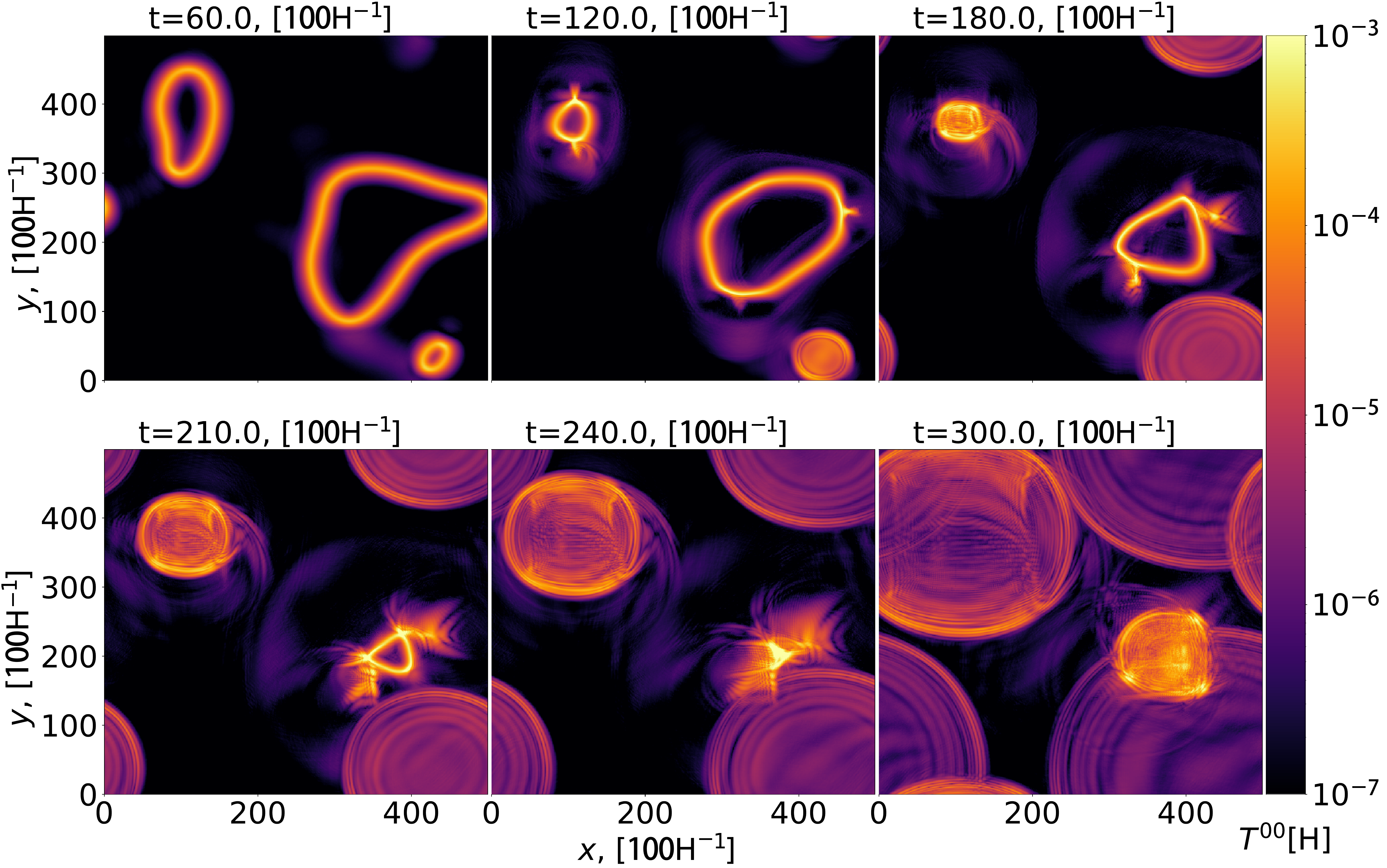}
  \caption{
   Collapse of closed domain walls (<<bubbles>>) as a result of surface tension. The xy-slice of the 3D simulation is shown. The color scale represents the energy density $\varepsilon\equiv T^{00}$ \eqref{edens} of field system $( \phi, \chi )$. During the collapse, the walls tend to get rid of curvature. After all, the energy is released by the scalar field radiation forming spherical waves (the bottom row). Energy density, distances and time are expressed in Hubble units $[H_\text{inf} = 10^{13}\,\text{GeV}]$.
  }
  \label{fig:Bubble_collapse}
  \vspace{-0.1cm}
\end{figure}

In the absence of significant curvature, domain walls are contracted due to the gradient terms in the field equation \eqref{fieldeq} and collapse with the energy release contained in the wall in the form of scalar field radiation. This effect was remarked in \cite{1989PhRvD..39.3576W}. Evolution of closed domain walls is demonstrated in Figure ~\ref{fig:Bubble_collapse}. The release of excess energy leads to the relaxation and spherization of the closed domain wall, which allows the wall to contract almost spherically.

Note, as a result of the closed domain walls collapse, the formation of primordial black holes is possible. This occurs when the mass of the domain wall is large enough to fit the wall within its gravitational radius during contraction. Such situation becomes possible under certain parameters of the field potential (see review \cite{2019EPJC...79..246B} and references within).

\subsection{Soliton foam}
\label{sec:foam}

\begin{figure}
  \centering
  \includegraphics[width=0.9\linewidth]{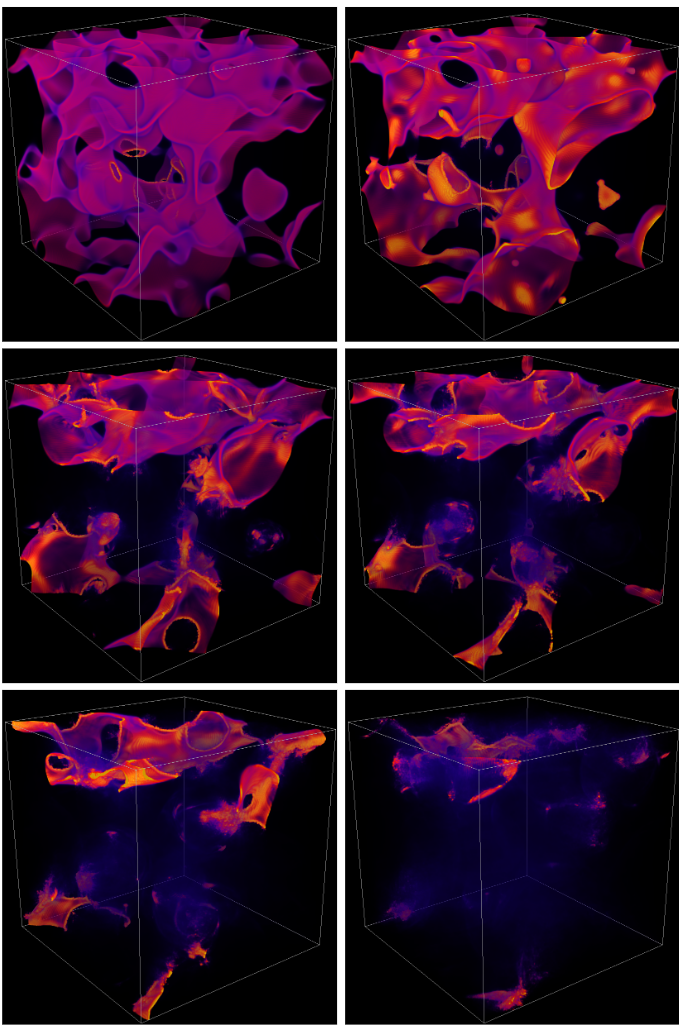}
  % \colorbox{black}{\includegraphics[trim={23mm 7mm 15mm 30mm}, width=0.42\linewidth]{foam1.pdf}}
  % \colorbox{black}{\includegraphics[trim={23mm 7mm 15mm 30mm}, width=0.42\linewidth]{foam2.pdf}}\vspace{1mm}
  % \colorbox{black}{\includegraphics[trim={23mm 7mm 15mm 30mm}, width=0.42\linewidth]{foam3.pdf}}
  % \colorbox{black}{\includegraphics[trim={23mm 7mm 15mm 30mm}, width=0.42\linewidth]{foam4.pdf}}\vspace{1mm}
  % \colorbox{black}{\includegraphics[trim={23mm 7mm 15mm 30mm}, width=0.42\linewidth]{foam5.pdf}}
  % \colorbox{black}{\includegraphics[trim={23mm 7mm 15mm 30mm}, width=0.42\linewidth]{foam6.pdf}}\vspace{1mm}

  \caption{Evolution of dense soliton foam formed by realistic inflationary initial conditions defined in section~\ref{sec:infcond}. Such dense configurations arise when the initial field values $\left(\phi_\text{in},\,\chi_\text{in}\right)$ are near the center of the saddle point of the potential \eqref{poten}. The colors show fields energy density distribution in 3-d space with warmer colors representing higher energy densities. Simulation size $L^3=900^3\cdot100^3\,[H_\text{inf}^{-3} ]$, simulation time step between frames $\Delta t\approx10^4\,[H_\text{inf}^{-1}]$.}
  \label{fig:foam}
\end{figure}

\begin{figure}
  \centering
  \includegraphics[width=0.9\linewidth]{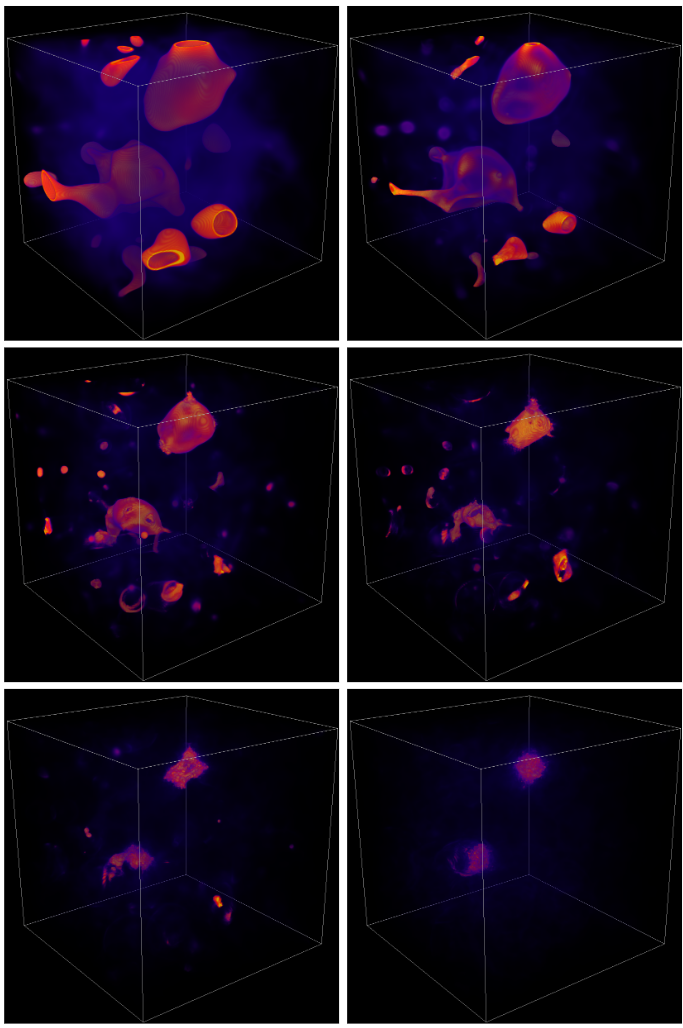}
  % \colorbox{black}{\includegraphics[trim={23mm 7mm 15mm 30mm}, width=0.42\linewidth]{buble1.pdf}}
  % \colorbox{black}{\includegraphics[trim={23mm 7mm 15mm 30mm}, width=0.42\linewidth]{buble2.pdf}}\vspace{1mm}
  % \colorbox{black}{\includegraphics[trim={23mm 7mm 15mm 30mm}, width=0.42\linewidth]{buble3.pdf}}
  % \colorbox{black}{\includegraphics[trim={23mm 7mm 15mm 30mm}, width=0.42\linewidth]{buble4.pdf}}\vspace{1mm}
  % \colorbox{black}{\includegraphics[trim={23mm 7mm 15mm 30mm}, width=0.42\linewidth]{buble5.pdf}}
  % \colorbox{black}{\includegraphics[trim={23mm 7mm 15mm 30mm}, width=0.42\linewidth]{buble6.pdf}}\vspace{1mm}

  \caption{Evolution of bubbles cluster formed by realistic inflationary initial conditions defined in section~\ref{sec:infcond}. Such configurations arise when the initial field values $\left(\phi_\text{in},\,\chi_\text{in}\right)$ are shifted away along direction $\phi$ (but still close) from the center of the saddle point of the potential \eqref{poten}. The colors show fields energy density distribution in 3-d space with warmer colors representing higher energy densities. Simulation size $L^3=900^3\cdot100^3\,[H_\text{inf}]$, simulation time step between frames $\Delta t\approx10^4\,[H_\text{inf}^{-1}]$.}
  \label{fig:bubble}
\end{figure}

In the case of real inflationary initial conditions (Figure~\ref{fig:init_fields}), the distribution of solitons in 3D physical space is found to be a foam-like. In this foam, the previously discussed effects appear simultaneously in different areas of the physical space. The structure of the soliton foam depends on the initial conditions described in section~\ref{sec:infcond}. Some random initial configurations form a dense sponge-like foam (Figure~\ref{fig:foam}) of interconnected domain walls with possible string-edged holes. Other random initial configurations evolve to a bubble-like foam (Figure~\ref{fig:bubble}) which is a more sparse cluster of closed domain walls.

\begin{figure}[H]
  \vspace{-0.1cm}
  \centering
  \includegraphics[width=1.0\linewidth]{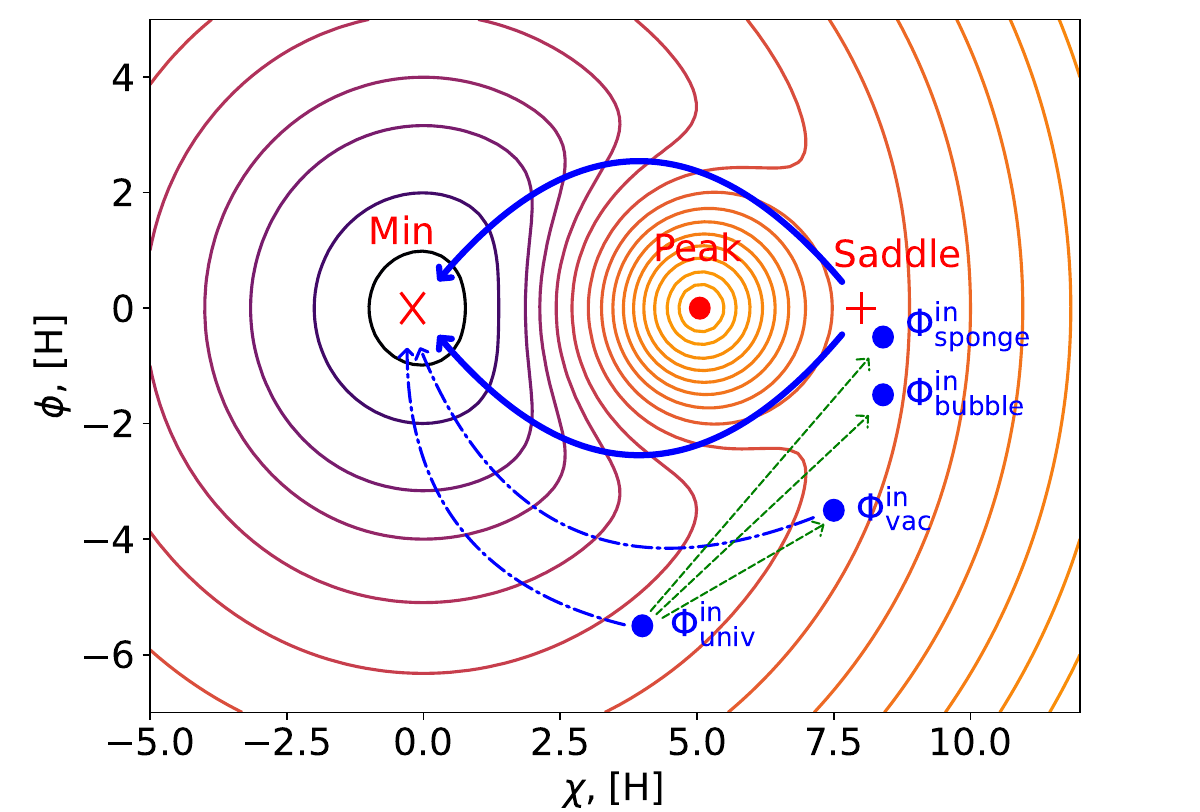}
  \caption{
   Locations of the initial field values $\Phi^\text{in}~\equiv~(\phi^\text{in}, \chi^\text{in})$ on field potential, leading to different (sponge-like, bubble-like and vacuum) scenarios. The contour lines illustrate the potential \eqref{poten}, its key points marked in red. The blue arrows show the different classical field evolutions which occurs after inflation. Green arrows show inflationary quantum fluctuations that shift the field from its initial value $\Phi^\text{in}_\text{univ}$, (initial value for the entire Universe) to the values $\Phi^\text{in}_\text{sponge}$, $\Phi^\text{in}_\text{bubble}$, forming soliton foam, or more distant values $\Phi^\text{in}_\text{vac}$, rolling straight into a vacuum.}
  \label{fig:scheme}
  \vspace{-0.1cm}
\end{figure}

One can note that these configurations are not isolated from each other, but instead represent the internal (sponge-like) and external (bubble-like) regions of one large soliton cluster (which cannot be modeled in this work due to its huge size). We can schematically illustrate this situation in Figure~\ref{fig:scheme}, where the location of the initial field values $\Phi^\text{in}~\equiv~(\phi^\text{in},\chi^\text{in})$ leading to both configurations are shown on the field potential. The initial values $\Phi^\text{in}_\text{sponge}$ which are close to the saddle point densely fill the space with the soliton foam, since the probability of rolling down on both sides of the peak substantially increases (it is precisely this rolling down that creates domain walls). If the initial values $\Phi^\text{in}_\text{vac}$ are far from the saddle point, the field evolution almost exclusively lead to the one-sided rolling, and solitons are almost not formed.

Further, one can note that physically close regions of space should also have close initial field values $\Phi^\text{in}$, since there is the spatial correlation described by the power spectrum $P(k)$ (see section \ref{sec:infcond}). Therefore, if we physically move away from the region filled with dense sponge-like foam (Figure~\ref{fig:foam}), we also move away from the point $\Phi^\text{in}_\text{sponge}$ in the field space, passing through the point $\Phi^\text{in}_\text{bubble}$. This last point, in turn, lead to the formation of region filled with sparse bubble-like foam (Figure~\ref{fig:bubble}). Likewise, if we move even further in the physical space, we get to points in the field space $\Phi^\text{in}_\text{vac}$ where solitons are not produced at all.

Thus, a soliton foam cluster should have a layered structure, in which the denser internal layers gradually thin out passing into vacuum. It is obvious that the total number of such soliton foam clusters in the Universe should be determined by the global initial field value at the start of inflation $\Phi^\text{in}_\text{univ}$, and the parameters of the cluster itself are closely related to the parameters of the potential \eqref{poten}. We discuss the cosmological implications of such soliton foam clusters in the next section.

\section{Discussion and conclusion}
\label{sec:discuss}

In this work, the non-thermal mechanism of the formation of complex soliton structures (<<soliton foam>>) is considered. This structures arise as a result of multiple quantum fluctuations of the scalar fields during inflation. We present the numerical 3D simulation of the soliton foam formation from the realistic initial field distribution and discuss its subsequent evolution. A field model is considered that has a potential \eqref{poten} with at least one local peak and a saddle point. This shape generalizes a wide class of potentials which includes, for instance, the well known <<Tilted Mexican Hat>> field model \cite{2018JCAP...04..042G}. Thus, the following conclusions might find an application in models other than those considered in this article.

We find that the soliton foam substantively consists of closed domain walls (bubbles), domain walls bounded by cosmic strings, and scalar field radiation (these basic structures have been previously discussed in literature \cite{1985PhR...121..263V}). Bubbles, domain walls bonded by strings and radiation interact with each other in a complex way leading to a dense sponge-like soliton foam (see Fig.~\ref{fig:foam}) in the center part of a soliton foam cluster.
The external regions of the soliton cluster has the bubble-like structure (see Fig.~\ref{fig:bubble}). Actually, a soliton foam cluster has a layered structure where the denser internal layers gradually thin out passing into vacuum from soliton sponge-like foam through separated bubbles to vacuum. Thus, the soliton foam tends to form locally and doesn't fill the entire Universe by contrast with the soliton networks scenario.

Moreover, for the studied class of potentials~\eqref{poten}, whether the soliton foam forms in the early Universe depends only on initial field values at the inflation epoch (see Fig.~\ref{fig:scheme}). We simulate the realistic initial conditions arising from multiple quantum fluctuations of inflationary epoch. The produced initial fields distributions allow to form local soliton networks (soliton foam clusters) without the need for the thermal production of a soliton network throughout the Universe. 

Potentially, the results of this paper can be extrapolated to arbitrary cosmological scales due to the scale invariance of inflationary quantum fluctuations which caused by the flat power spectrum $P(k)$. In this case, the soliton density varies depending on the initial conditions $\Phi^\text{in}$ for different scales.

As we shown in section \ref{sec:results}, the evolution of soliton structures is accompanied by significant radiation emission. Far from the soliton structures, this emission could be interpreted as the dark matter \cite{2012PhRvD..85j5020H, 2014JCAP...03..037H}. Moreover, the collapse of soliton structures could lead to primordial black holes formation through the several mechanisms \cite{1989PhLB..231..237H, 1993PhRvD..47.3265G, 2000hep.ph....5271R, 2000IJMPA..15.4433H, 2001JETP...92..921R, 2019PhRvD..99j4028H, 2019EPJC...79..246B, 2020PhRvD.101b3513L}. In turn, PBHs itself have been actively considered as dark matter candidates \cite{2020ARNPS..70..355C}; this approach has its pros and cons \cite{2021RPPh...84k6902C}. 

Taking into account PBHs creating mechanism based on domain wall collapse, one can note that one model \eqref{action} can create dark matter consisting of three components with the different spatial distribution: a) diffuse homogeneous background of coherent oscillations of the scalar fields formed after the inflation; b) halos of massive scalar particles formed by radiation emission as a result of soliton foam clusters relaxation and decay; c) compact PBH clusters formed as a result of the collapse of closed domain walls. Thus, the model might naturally predict both the appearance of primordial black hole seeds in the early Universe, and dark matter halos around them, and a diffuse homogeneous dark matter component, which is the promising opportunity.

Estimating of all dark matter components produced in this model will be the subject of further research aimed at searching for cosmological distributions of various components of the soliton foam (in this work we focus on the local properties of the soliton structures). In general, this search comes down to determining how the spatial density of the basic components of a soliton foam depends on the initial field values in a given region of space formed during cosmological inflation. This problem could be solved by a multiple numerical experiments (similar to the simulations shown in Figures~\ref{fig:foam}--\ref{fig:bubble}) to study the entire region of interest of initial field values (Figure~\ref{fig:scheme}). Furthermore, by calculating the scale distributions of inflationary quantum fluctuations (forming these initial values), we can predict the size spectrum of the formed soliton foam clusters in the Universe, as well as the total contribution of the various soliton foam components to the critical density $\rho_c$. In addition, the discussed approach allows to test the influence of formed soliton clusters on the cosmic microwave background. We do not expect substantial deviations from the observational data. However, these physical implications are the question of a separate investigation.

\section*{Acknowledgements}
We are grateful to S.~G.~Rubin, K.~M.~Belotsky, M.~Y.~Khlopov and V.~D.~Stasenko for useful discussions and their interest in the work.
The work was funded by the Ministry of Science and Higher Education of the Russian Federation, Project <<New Phenomena in Particle Physics and the Early Universe>> FSWU-2023-0073.

\sloppy

\printbibliography

\end{document}